\documentclass[aps,prb,twocolumn,superscriptaddress,10pt]{revtex4}

\usepackage{graphicx}
\usepackage{amsmath}
\usepackage{amssymb}
\usepackage{amsfonts}
\usepackage{bm}
\usepackage{psfrag}
\usepackage{pstricks}
\usepackage{multirow}

\usepackage{color}

\begin{document}
\title{Interface roughness, carrier localization and wave function overlap in $c$-plane InGaN/GaN quantum wells: Interplay of well width, alloy microstructure, structural inhomogeneities and Coulomb effects}

\author{Daniel S. P. Tanner}
\affiliation{Photonics Theory Group, Tyndall National Institute,
University College Cork, Cork, T12 R5CP, Ireland}
\author{Joshua M. McMahon}
\affiliation{Photonics Theory Group, Tyndall National Institute,
University College Cork, Cork, T12 R5CP, Ireland}
\author{Stefan Schulz}
\affiliation{Photonics Theory Group, Tyndall National Institute,
University College Cork, Cork, T12 R5CP, Ireland} 

\begin{abstract}
In this work we present a detailed analysis of the interplay of
Coulomb effects and different mechanisms that can lead to carrier
localization effects in $c$-plane InGaN/GaN quantum wells. As
mechanisms for carrier localization we consider here effects
introduced by random alloy fluctuations as well as structural
inhomogeneities such as well width fluctuations. Special attention
is paid to the impact of the well width on the results. All
calculations have been carried out in the framework of atomistic
tight-binding theory. Our theoretical investigations show that
independent of the  here studied well widths, carrier localization
effects due to built-in fields, well width fluctuations and random
alloy fluctuations dominate over Coulomb effects in terms of charge
density redistributions. However, the situation is less clear cut
when the well width fluctuations are absent. For large well width
(approx. $>$ 2.5 nm) charge density redistributions are possible but
the electronic and optical properties are basically dominated by the
spatial out-of plane carrier separation originating from the
electrostatic built-in field. The situation changes for lower well
width ($<$ 2.5 nm) where the Coulomb effect can lead to significant
charge density redistributions and thus might compensate a large
fraction of the spatial in-plane wave function separation observed
in a single-particle picture. Given that this in-plane separation
has been regarded as one of the main drivers behind the green gap
problem, our calculations indicate that radiative recombination
rates might significantly benefit from a reduced quantum well
barrier interface roughness.
\end{abstract}

\date{\today}

\maketitle

\section{Introduction}

Over the last several years, the theoretical and experimental
analysis of carrier localization effects in InGaN/GaN quantum wells
(QWs) has been of strong
interest,~\cite{ChGa1998,MoLe02,GrSo05,ChUe06} and has recently
gathered enormous
pace.~\cite{WaGo2011,HaWa12,YaSh2014,ScCa2015,DaSc16,TaCa2016_RSC,MaPe16,JoTe2017,Karpov_17,FiPi2017,LiPi2017,PiLi2017}
This stems from the fact that a detailed understanding of this
question is important not only  from a fundamental physics point of
view, but also for device applications.~\cite{ChUe06} Indeed,
carrier localization is widely accepted to be the reason that modern
light emitting devices based on InGaN/GaN QWs, with their high
density of defects, are able to function at all.~\cite{ChUe06}
Furthermore, carrier localization is essential to explain the
optical properties of this material system in detail. For example,
the photoluminescence (PL) spectra of these InGaN/GaN QWs exhibit
large line widths,~\cite{GrSo05,SoEs2015} an ``S-shaped''
temperature dependence of the PL peak energy,~\cite{ChGa1998,HaWa12}
a mobility edge,~\cite{BlSc2018} and a time-decay behavior where the
explanation depends crucially on localization
effects.~\cite{MoLe02,DaSc16}

The cause, nature, and consequence of this localization has
generated significant debate over the years.
Initially localization in InGaN was attributed to observed In rich
clusters\cite{Chichibu_Cluster} purportedly forming due to the
theoretically determined immiscibility of the alloy.~\cite{HoSt96}
However, careful transmission electron microscopy
(TEM)~\cite{SmKa2003} and detailed atomic force microscopy (AFM)
measurements~\cite{GaOl07} found the observed clusters to be
artifacts of the measurement technique. Additionally, when studies
accounted for the strain in InGaN QWs, originating from an
underlying substrate, it was demonstrated that the miscibility limit
is higher than predicted from a calculation where this aspect is
neglected.~\cite{Karpov_1998} Since these developments, localization
in $c$-plane InGaN QWs has been primarily attributed to random alloy
fluctuations~\cite{WaGo2011,ScCa2015,TaCa2016_RSC,MaPe16,LiPi2017,JoTe2017}
along with features introduced by the interplay of built-in field
and well width fluctuations
(WWFs),~\cite{WaGo2011,ScCa2015,TaCa2016_RSC} which have been
observed in several experimental studies.~\cite{GrSo05,SaHe14}
The relationship between these mechanisms of localization and the
manner in which the electrons and holes are respectively localized,
and the consequences of this localization, stand as the next
challenge to the full understanding and optimization of InGaN
QW-based light emitting devices. This originates from the fact that
recently two of the major roadblocks for future energy efficient
light emission from InGaN-based LEDs, namely the ``green gap'' and
the ``droop'' phenomenon, have been tightly linked to carrier
localization features and interfacial
roughness.~\cite{YaSh2014,MaPe16,Karpov_17,LiPi2017,JoTe2017,TaSu2017}

For instance, interfacial roughness between an InGaN QW and the GaN
barrier has been considered a key factor for increasing Auger
recombination rates, as shown by Tan \emph{et al.}~\cite{TaSu2017}
Several studies have reported that Auger recombination is the main
driver behind the efficiency droop problem, which describes the
reduction in the efficiency of InGaN-based LEDs with increasing
current density.~\cite{KiRi11,BiGa13,IvWe13,GaSa13} In a separate
theoretical study, which neglected WWFs, Jones \emph{et
al.}~\cite{JoTe2017} demonstrated that compared to virtual crystal
approximation (VCA) calculations, calculations including alloy
induced carrier localization features reveal increased Auger
recombination rates in InGaN QW systems. Both the work of Tan
\emph{et al.}~\cite{TaSu2017} and Jones \emph{et
al.}~\cite{JoTe2017} build on the same underlying concept that
$k$-selection rules are broken due to alloy/interfacial roughness
induced localization. This highlights that understanding the effect
of WWFs and alloy fluctuations on the electronic and optical
properties of $c$-plane InGaN QWs in detail is not only of interest
from a fundamental perspective but also from a device application
point of view.

The understanding of these features is also relevant to the green
gap problem. This describes the drop in efficiency of InGaN/GaN
emitters in the green to yellow spectral range when compared to blue
emitters. Auf der Maur \emph{et al.}~\cite{MaPe16} analyzed the
contribution of random alloy effects to the green gap phenomenon.
The authors~\cite{MaPe16} draw the conclusion that not only does the
built-in field serve to decrease the radiative recombination rate,
but so too does the spatially in-plane separation of the carriers
due to localization features induced by the random alloy. Using as a
reference a calculation based on a VCA, the authors concluded that
this in-plane separation contributed as much as 30\% to the
reduction of the radiative recombination rate in green-emitting
QWs.~\cite{MaPe16} A similar conclusion was reached by Karpov using
an empirical model.~\cite{Karpov_17}

However, it should be noted that a recent investigation by Jones
\emph{et al.}\cite{JoTe2017} contradicted this result. Using a
modified continuum-based approach, accounting for random alloy
fluctuations and connected carrier localization effects, Jones
\emph{et al.}~\cite{JoTe2017} showed that random alloy fluctuations
are beneficial for the radiative recombination rate. This is based,
again, on the argument that with random alloy fluctuations,
$k$-selection rules for transitions between different electron and
hole states are no longer valid. Thus, even though the dipole matrix
elements are reduced by carrier localization effects and the
connected spatial separation of electron and hole wave functions, a
significantly increased number of transitions are allowed when
compared to a VCA treatment. So, taking all this into account, the
impact of carrier localization effects on the optical properties and
in particular on the question of the green gap problem, which is
crucial for device applications, is still surrounded by controversy
as is the underlying physics of these systems in general.

In addition to their disagreement on the impact of alloy
fluctuations on the radiative recombination in InGaN devices, these
previous theoretical studies have also neglected two important
aspects of typical InGaN/GaN QWs. First, the above discussed studies
do not account for structural inhomogeneities such as WWFs. These
structural features, in tandem with the strong built-in fields in
$c$-plane InGaN QWs, have been found to localize electron wave
functions,~\cite{Wats2011,ScCa2015} and can therefore strongly
contribute to the in-plane carrier separation. If the in-plane
carrier localization features are central to the green gap problem,
then WWFs, by inducing further in-plane separation, should worsen
the reduction in efficiency. Therefore, a detailed understanding of
the impact of the interface roughness is required for the
optimization of devices to ameliorate the green gap problem.

Second, an additional component when studying the importance of
in-plane carrier localization for optical properties are Coulomb
effects. These effects have been widely neglected in the theoretical
studies that account for carrier localization features in InGaN/GaN
QWs. This leads to the question of whether or not Coulomb effects
can reduce the spatial in-plane carrier separation. Given that the
electrostatic built-in field strongly spatially separates electron
and hole wave functions along the growth direction, the importance
of Coulomb effects should increase with decreasing well width. In
this respect, the question remains if carrier localization due the
combination of WWFs, built-in field and random alloy fluctuations
dominates over the attractive Coulomb interaction between electron
and hole. Most of the above discussed theoretical studies have been
either based on single-particle
results,~\cite{JoTe2017,MaPe16,WaGo2011,TaCa2016_RSC,FiPi2017}, in
the absence of WWFs, or for a fixed well
width.~\cite{JoTe2017,MaPe16,FiPi2017} In studies including Coulomb
effects and carrier localization, a fixed well width and a high In
content of 25\% has been targeted so far.~\cite{ScCa2015}

Finally, if Coulomb effects do indeed gain significance when
structural well parameters change, the recombination dynamics of the
system could change drastically, which again could be of benefit for
InGaN-based devices with improved radiative recombination rates. An
important early connection between carrier localization mechanisms
and the distinctive optical properties of InGaN/GaN QWs was
demonstrated by Morel \emph{et al.}~\cite{MoLe02} Here, Morel and
co-workers~\cite{MoLe02} applied a model of independently localized
electrons and holes (pseudo two-dimensional donor-acceptor pair
system) to describe the \emph{non-exponential} decay curves in
time-resolved PL measurements of $c$-plane InGaN/GaN QW systems. In
contrast, other studies on $c$-plane InGaN/GaN QWs reported
time-resolved PL measurements in which the PL decay curves exhibit a
\emph{single exponential} behavior.~\cite{DaDa00} Such behavior can
be attributed to exciton localization effects, where, for instance,
the hole is strongly localized and the electron localizes about the
hole.~\cite{DaSc16} Therefore, as already indicated above, the
recombination dynamics (non-exponential vs. single exponential) can
be affected and ideally may be tuned by a variety of different
factors such as well width,~\cite{DaDa00} In
content,~\cite{MoBe99,NaYu97} or as we will show, the interface
roughness. Taking all of the above into consideration, understanding
the interplay of Coulomb effects, interfacial roughness and carrier
localization mechanisms is obviously interesting from a fundamental
physics view point but is also key for device design.

In this work we undertake the analysis of the interplay of carrier
localization mechanisms, introduced by (random) alloy fluctuations,
WWFs and Coulomb effects and we comment here on the connected impact
on the optical (radiative) properties of InGaN/GaN QWs. Special
attention is paid to the influence of the well width on the results.
The calculations have been carried in an atomistic tight-binding
(TB) framework, allowing us to gain insight into these questions on
a microscopic level.

Here, we have studied $c$-plane InGaN/GaN QWs with a well width
varying between 1.6 nm and 3.4 nm. For this analysis we have used an
``intermediate'' In content of 15\%. By studying identical QWs with
and without WWFs, we demonstrate the significant impact that WWFs
have on the localization of electron states, even for smaller well
width (1.6 nm). Independent of well width, this picture is mainly
unchanged when Coulomb effects are included and carrier localization
features remain dominated by WWFs (electron) and random alloy
fluctuations (hole). For higher In contents, usually realized for
emitters operating in the green spectral range, this effect will be
even more pronounced.~\cite{TaCa2016_RSC}

The situation in the absence of WWFs is less clear cut. Here, the
single-particle electron wave function exhibits a more delocalized
nature with perturbations introduced by the local alloy structure.
When including Coulomb effects, even for the larger well width, our
calculations exhibit a redistribution of the electron charge
density. When turning to systems with a lower well width, Coulomb
effects can significantly affect the spatial in-plane carrier
separation. Here, we observe that in the absence of the WWFs the
electron wave function is much more likely to localize about the
hole, thus resulting in exciton localization-like features. Overall,
when looking at optical spectra, the benefits of this charge density
redistribution are found to increase with decreasing width. This
indicates that the out-of plane separation of the carriers dominates
the wave function overlap at larger widths.

In general, our results show that reducing the interface roughness
between GaN and InGaN in InGaN/GaN QWs, and thus ideally
circumventing WWFs, should, due to Coulomb effects, reduce the
in-plane spatial separation of electron and hole wave functions.
Following the idea of Auf der Maur~\emph{et al.}~\cite{MaPe16}, the
radiative recombination rate should benefit from this. Additionally,
in the light of the arguments given by Jones~\emph{et
al.}~\cite{JoTe2017}, an increased wave function overlap along with
the absence of $k$-selection rules should give a further improvement
of the radiative recombination rate. Thus reducing the InGaN/GaN
interface roughness could be a promising way forward to improve the
radiative device characteristics of InGaN/GaN based light emitters
across a wide range of emission wavelength.

The manuscript is organized as follows. In the following section we
briefly review the theoretical framework used in our studies. Here,
also the QW system and how WWFs and alloy microstructure are treated
is discussed. Section~\ref{sec:Results} presents the results of our
theoretical study. In a first step,
Sec.~\ref{sec:Built_in_potential}, the impact of random alloy
fluctuations and WWFs on the built-in potential are discussed. This
is followed by the analysis of single-particle results in
Sec.~\ref{sec:electronic_structure_SP}. The impact of Coulomb
effects on optical properties is discussed in
Sec.~\ref{sec:electronic_structure_MB}. In Sec.~\ref{sec:comp_exp},
we compare and relate our theoretical findings to experimental data,
radiative recombination dynamics, and the connected importance for
nitride-based optoelectronic devices. Finally, in
Sec.~\ref{sec:Conclusion} a summary and conclusions of our
investigations are presented.

\section{Theory and QW system}
\label{sec:Theory}

In this section we briefly review the theoretical framework and the
QW model system applied in this work. We start in
Sec.~\ref{sec:Theory} with the theoretical framework. In
Sec.~\ref{sec:QW_structures} the supercell used in our calculations
is introduced. Here, we discuss also aspects such as In atom
distribution as well as structural inhomogeneities (well width
fluctuations).

\subsection{Theoretical Framework}
\label{sec:Theory}

The full description of each of the ingredients of our theoretical
framework has been presented in detail in our previous
work.~\cite{ScCa2015} Here, we summarize only the main points. The
central component underlying our study of the electronic and optical
properties of InGaN/GaN QWs on an atomistic level is a nearest
neighbor $sp^3$ tight-binding (TB) model. This model takes input
from a valence force field (VFF) model, implemented in
LAMMPS.~\cite{LAMMPS} The applied VFF approach is based on Martin's
model, introduced in Ref.~\onlinecite{Ma1970}, including therefore
electrostatic effects. This model accurately captures deviations
from the ideal wurtzite structure, such as the lattice constant
$c/a$ ratio. Given the atomistic nature of our VFF approach, the
model takes into account that the number of In and Ga atoms in the
nearest neighbor environment of a N atom varies. Using the VFF, the
relaxed atomic positions in the alloyed and strained supercell can
be determined, including therefore also alloy induced local
variations in the bond length and corresponding strain fields. These
relaxed atomic positions provide input for the strain corrections of
the TB matrix elements.~\cite{CaSc2013} Additionally, the relaxed
atomic positions provide input for our recently developed local
polarization theory.~\cite{CaSc2013} In this approach, the total
polarization vector field is divided into a macroscopic and local
polarization contribution. The macroscopic part is related to
clamped ion contributions (no relaxation of the internal degrees of
freedom of the atoms), while the local part is evaluated per local
tetrahedron accounting therefore for internal strain effects.
Starting from the local polarization at the different lattice sites,
the resulting built-in potential is calculated from a point
dipole-model. This circumvents the problem of solving Poisson's
equation on a (non-uniform) wurtzite lattice, keeping in mind the
relaxed atomic positions of the InGaN QW. The calculated built-in
potential is included in the TB model as a site-diagonal correction,
which is a widely used
approximation.~\cite{Saito2002,ZiJa2005,ScBa2012} It should be noted
that we are here interested in the low carrier density regime (see
below). Focussing on this regime helps to disentangle carrier
localization effects arising from WWFs, alloy microstructure and
Coulomb effects, thus throwing light on the interplay of these
different aspects. Therefore, a non-self-consistent TB calculation
is sufficient for our purposes here. However, to account for
potential charge carrier density redistributions, which could be
important, for instance, for the green gap problem, as highlighted
above in more detail, we include Coulomb effects in our
calculations.

To include Coulomb effects, the TB single-particle states serve as
input for configuration interaction calculations.~\cite{BaSc2007}
The calculations are restricted to excitonic effects, meaning that
only one electron-hole pair is considered. We account here for the
direct electron-hole interaction; exchange terms are neglected since
the connected matrix elements are small compared to the direct part.
To account for screening effects in the Coulomb interaction, which
is intrinsically a complex problem,~\cite{LeLo1982,FrFu1999} we use
here a simplified approach and assume an isotropic and material
independent (static low frequency) dielectric constant. For GaN we
employ the values from Nakamura and Chichibu
($\epsilon^{r}_{\perp,0}=7.87$, $\epsilon^{r}_{\parallel,0}=8.57$),
resulting in a very good agreement between theory and experiment in
terms of GaN exciton binding energies.~\cite{NaCh2000} For InN we
use the values from Ref.~\onlinecite{DaKl1999}
($\epsilon^{r}_{\perp,0}=13.1$, $\epsilon^{r}_{\parallel,0}=14.4$).
Given that both for InN and GaN the vales of the ordinary
($\epsilon^{r}_{\perp,0}$) and extraordinary
($\epsilon^{r}_{\parallel,0}$) component are very similar in
magnitude, the isotropic approximation is reasonable. To obtain the
InGaN dielectric constant, a linear interpolation between the InN
and GaN values is applied. For the Coulomb calculations, this
averaged dielectric constant is used for the entire supercell,
justified by the observation that electron and hole wave functions
are mainly localized inside the QW region (see below). Furthermore,
as we will discuss below, we will study here only moderate In
contents so that the dielectric constant contrast between well and
barrier is small. Thus, the assumption of a position independent
dielectric constant is a reasonable first approximation. Given that
the screening of the Coulomb interaction should be distance
dependent, our Coulomb matrix elements are split into a short-range
and long-range part.~\cite{SwZi2017} The short-range, on-site part
is unscreened while for the long-range part the above approximation
for the screening has been used. Similar approaches and
approximations have been made in other
systems.~\cite{ShCh2005,GoBa2013,SwZi2017} In addition to Coulomb
matrix elements, the TB wave functions have also been used to
calculate dipole matrix elements.~\cite{ScSc2006,ScCa2015} In
connection with many body wave functions and via Fermi's golden
rule, the dipole matrix elements are used to calculate optical
spectra. More details are given in Refs.~\onlinecite{ScSc2006}
and~\onlinecite{ScCa2015}.

\subsection{QW structure and simulation supercell}
\label{sec:QW_structures}

The TB calculations have been performed on wurtzite supercells with
approximately 82,000 atoms. This corresponds to an overall system
size of 10$~$nm x 9$~$nm x 10$~$nm with periodic boundary
conditions. The QW width inside the supercell has been varied to
investigate the impact of this quantity on the electronic and
optical properties of the system. The well width, $L_{w}$, is set
to: $L_{w}=1.6~$nm, $L_{w}=2.11~$nm, $L_{w}=2.65~$nm and
$L_{w}$=3.43$~$nm. As mentioned above, we are interested in
investigating the influence of the well width on the electronic and
optical properties of InGaN/GaN QWs; to do so, we keep the In
content in the QW region fixed at 15\%. Based on experimental
studies,~\cite{GaOl2007,SmKa2003} we distribute In atoms randomly in
the active region of the QW. It is important note that in our
atomistic framework, the In content at the QW barrier interface will
locally vary. But, we do not account here for penetration of In
atoms into the barrier. In the literature different In atom
distribution profiles have been considered. Yang \emph{et
al.}~\cite{YaSh2014} and McBride \emph{et al.}~\cite{McBYa2014}
considered a Gaussian alloy distribution along the growth direction
of $c$-plane InGaN QWs. However, several experimental studies have
reported a sharp lower QW barrier interface (growth of InGaN on
GaN), while the upper interface (GaN on InGaN) exhibited WWFs and
some penetration of In into the
barrier.~\cite{MoBe1999,GaOl2007,HoHa2013,OlBe2010,SaNa2017,MaPi2017}
Watson-Parris~\cite{Wats2011} considered the effect of In
incorporation in the barrier at the upper interface for a fixed well
width (2.85 nm) and In content (25 \%) on the basis of a modified
continuum-based model, concluding that the inclusion of a diffuse
upper interface has no noticeable effect on the results in terms of
carrier localization. This indicates that for our present study the
penetration of In atoms into the barrier is of secondary importance.
This is further supported by the fact that the diffuseness of the In
atoms in the QW plane is homogenous.~\cite{WaGo2011} Thus, one could
expect that the experimentally observed WWFs have a more significant
impact on carrier localization effects when compared to the
penetration of In atoms into the barrier. Finally, it should be
noted that recent experimental studies have shown that by a careful
choice of the growth procedure, the In incorporation in the barrier
can significantly be reduced.~\cite{MaPi2017} Therefore, the here
made assumptions about the In distribution in the well and at the QW
barrier interface should provide a good description of the
experimentally observed systems.

In order to analyze the impact of the alloy microstructure on the
results, the calculations have been repeated 20 times for each $L_w$
value. To closely compare the results from the same configuration
for different QW widths $L_{w}$, we proceed in the following way. We
start with the lowest $L_{w}$ values and generate the 20 different
microscopic configurations. When studying the next largest system,
for each configuration, we keep the random In atom configuration and
just add new layers to the existing QW structure. For instance, for
configuration (Config) 1 of a QW of width $L_{w}=2.11$~nm, the
placement of In atoms in the first 12 atomic layers are identical to
the placement of In atoms in the first 12 layers of the
$L_{w}=1.6$~nm system for the same configuration. Likewise the first
20 atomic layers of Config $i$ of the $L_{w}=2.85$ QW have identical
In atom distributions to the 20 atomic layers in Config $i$ forming
the well with $L_{w}=2.11$~nm. Thus, in terms of the In atom
distribution, the lower QW-barrier interface is always the same for
a given microscopic configuration, independent of the QW well width.
In this way we are able to partially isolate the effects of well
width and WWFs on the electronic and optical properties of the here
considered In$_{0.15}$Ga$_{0.75}$N/GaN QWs from random alloy
effects. Finally, all calculations have been performed in the
presence and absence of WWFs. For the WWFs, following previous
work,~\cite{WaGo2011,Wats2011,ScCa2015,TaCa2016_RSC} we assume
disk-like WWFs. The diameter of these fluctuations has been assumed
to be 5 nm with a height of 2 monolayers, which is consistent with
experimental observations.~\cite{GrSo05,SaHe14} In total 320
atomistic calculations have been performed to study the interplay of
Coulomb effects, well width and carrier localization due to WWFs and
random alloy fluctuations on the electronic and optical properties
of In$_{0.15}$Ga$_{0.85}$N/GaN QWs.

\begin{figure*}[t!]
\includegraphics{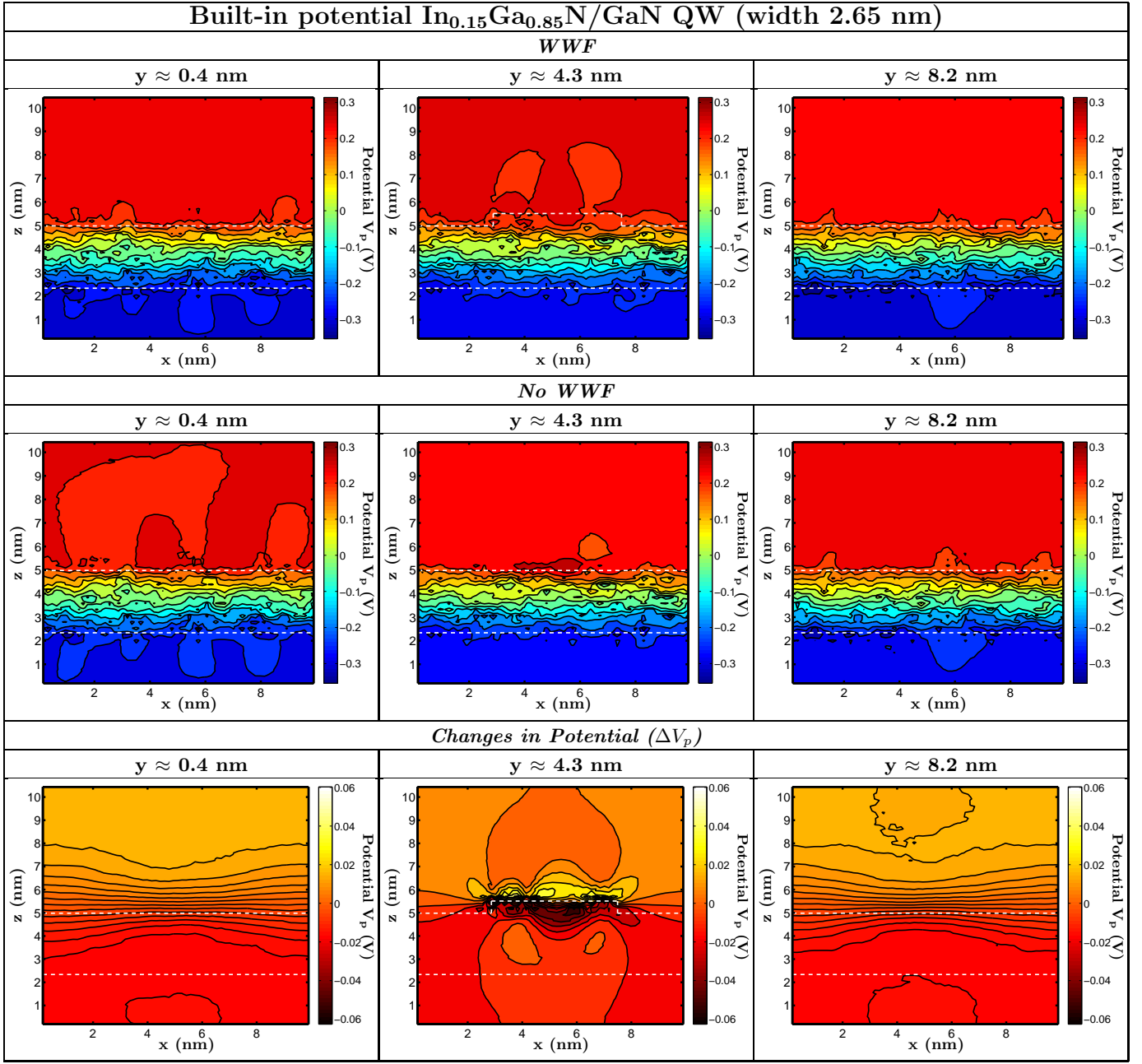}
\caption{(Color online) Contour plots of the electrostatic built-in
potential $V_p$ in the $x$-$z$-plane of an
In$_{0.15}$Ga$_{0.85}$N/GaN quantum well with well width $L_w=2.65$
nm. The $z$-axis is parallel to the wurtzite $c$-axis. The results
are shown for different slices ($y=0.4$ nm; $y=4.3$ nm; $y=8.3$ nm)
through the supercell of the same arbitrary microscopic
configuration. The first row depicts the data for the quantum well
where a disk-like well width fluctuation has been taken into account
(\emph{WWF}). The middle row displays the results for the same
configuration and slices through the same planes as in the first
row, however, this time in the absence of the well width fluctuation
(\emph{No WWF}). The last row depicts the difference in the
potential $\Delta V_p=V^\text{WWF}_p- V^\text{No WWF}_p$ between the
situation with ($V^\text{WWF}_p$) and without ($V^\text{No WWF}_p$)
the well width fluctuation calculated at the different
$y$-coordinates.} \label{fig:slice_phi_p}
\end{figure*}

\section{Results}
\label{sec:Results}

In this section we present the results of our atomistic analysis of
the impact of well width and WWFs on the electronic and optical
properties of $c$-plane InGaN/GaN QWs. Before turning to the
electronic and optical properties, we start our discussion with a
study of the electrostatic built-in potential and how alloy and WWFs
affect this quantity. This investigation is presented in
Sec.~\ref{sec:Built_in_potential}. Having discussed the built-in
potential, we turn and present in
Sec.~\ref{sec:electronic_structure_SP} the results of our TB
analysis in the absence of Coulomb effects (single-particle data).
In Sec.~\ref{sec:electronic_structure_MB} we discuss the impact of
Coulomb (excitonic) effects on the results. Finally, we relate the
here obtained theoretical results to experimental observations
(Sec.~\ref{sec:comp_exp}).

\subsection{Impact of random alloy and well width fluctuations on the built-in potential}
\label{sec:Built_in_potential}

To establish the impact of alloy and WWFs on the electrostatic
built-in potential, $V_p$, Fig.~\ref{fig:slice_phi_p} displays
contour plots of $V_{p}$ for different slices through the supercell
of the same arbitrarily chosen microscopic configuration. The
potential displayed is that of a QW of width $L_{w}=2.65$ nm. The
contour plots are shown in the $x-z$-plane, where $z$ is parallel to
the wurtzite $c$-axis. The $y$-coordinate is varied, with increasing
magnitude from left to right, resulting then in the different slices
depicted in subsequent columns of Fig.~\ref{fig:slice_phi_p}. As a
guide to eye, the (white) dashed lines indicate the QW interfaces.
To ascertain the impact of interface roughness on the built-in
potential: the upper row displays the result for a QW with a WWF
($V^\text{WWF}_p$); the middle row gives the same QW configuration
without a WWF ($V^\text{NoWWF}_p$); and the last row shows the
difference in the potential between a QW with and without a WWF,
$\Delta V_p=V^\text{WWF}_p-V^\text{NoWWF}_p$. Several features of
the built-in potential are now of interest for our analysis.

\begin{figure}
\includegraphics{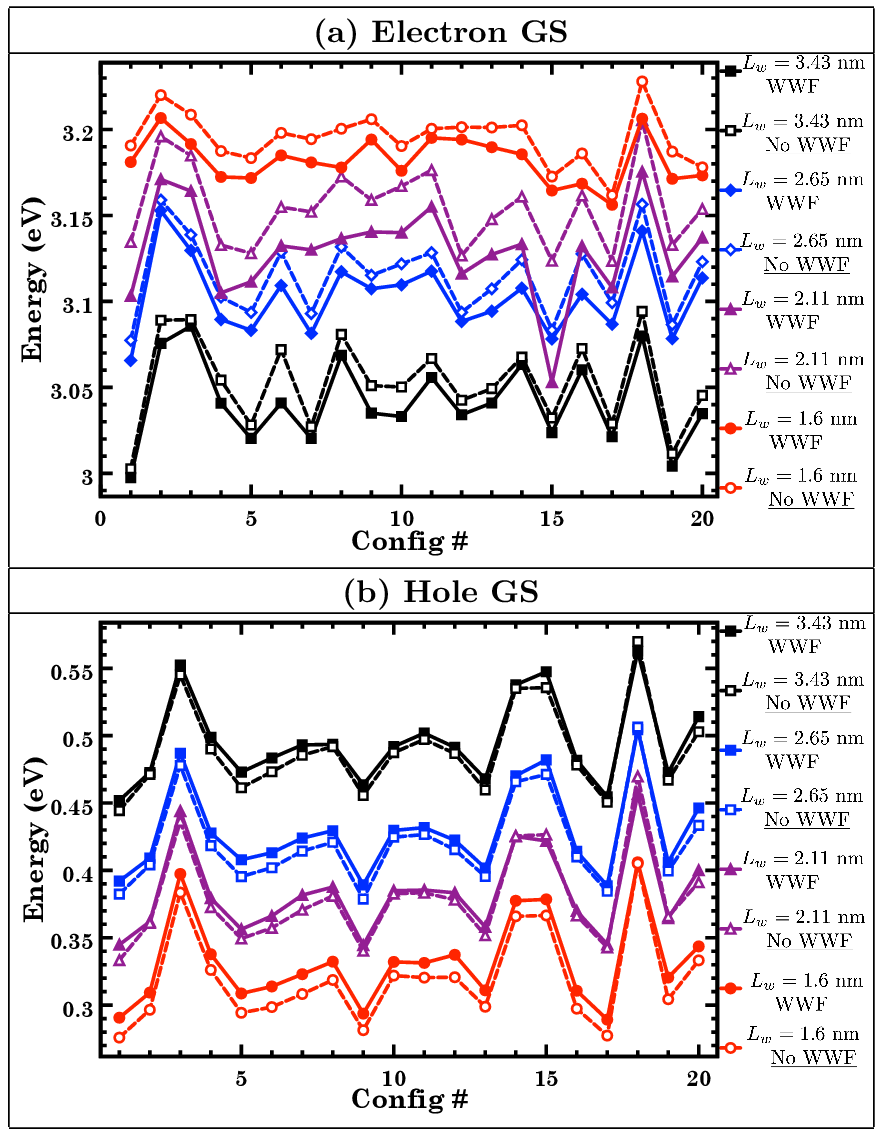}
\caption{Single-particle (a) electron and (b) hole ground state (GS)
energy as a function of the microscopic configuration number (Config
\#). The data from calculations including well width fluctuations
(WWF) are given by the filled symbols; the results when the well
width fluctuation is absent are denoted by the open symbols (No
WWF). The data are displayed for the different well widths $L_w$.}
\label{fig:SP_GS_electron_hole}
\end{figure}

First, in contrast to standard 1-D continuum-based approximations,
which treat InGaN/GaN QW systems as an ideal system that can be
described by average parameters, the built-in potential is strongly
position dependent. This position dependence is manifested in
Fig.~\ref{fig:slice_phi_p} as isolines which are non-parallel. This
is in contrast to the parallel isolines expected from a 1-D
continuum-based, capacitor-like picture, where the potential is
constant outside the well and has a well defined slope inside. Thus,
the corresponding field is zero in the barrier and constant inside
the QW. Second, the alloy fluctuations lead also to ``pockets'' in
which the built-in potential significantly varies with respect to
its environment. From this we infer already that random alloy
fluctuations lead to significant changes in the built-in potential,
both on a macroscopic as well as on a local, microscopic level.
Third, turning now to the impact of the WWF, as expected, in the
region of the WWF the built-in potential changes compared to the
situation without the WWF. However, when looking at the last row of
Fig.~\ref{fig:slice_phi_p}, we find also that the built-in potential
underneath the WWF changes significantly compared to the situation
in which no WWF has been considered. This originates from the fact
that the disk-like WWF is a three dimensional object and that the
strain field around this ``QD-like'' structure is changed. This has
several consequences. Given that electron wave functions are
localized at the upper interface of the QW and that the (alloy)
microstructure of the upper layer with the inclusion or exclusion of
a WWF changes, carrier localization features of these states are
expected to be strongly affected. What might be less straightforward
to predict \emph{a priori} is that even though the alloy
microstructure at the lower interface is not changed, the hole
states which are localized at this interface, can be affected by the
three-dimensional built-in potential effects discussed above. This
aspect will be more pronounced in narrower wells. Therefore,
introducing WWFs affects not only the effective volume of the QW,
but also produces built-in potential modifications originating from
the altered microscopic alloy configuration and strain relaxation of
the system.
Overall, the inclusion of the WWF also leads to a slight increase in
the macroscopic built-in field compared to the system without WWFs.
This is based on the observation of slightly negative (positive)
built-in potential values below (above) the well, as it can be
inferred from $\Delta V_p$ (cf. last row of
Fig.~\ref{fig:slice_phi_p}). Equipped with this knowledge, we turn
now to analyze the electronic structure of InGaN/GaN QWs in the
presence and absence of WWFs. Special attention is paid to the
impact of the well width $L_w$ on these properties. This will be the
topic of the next two sections. We begin our analysis in the next
section with an examination of single-particle results.

\subsection{Impact of well width, random alloy and well width fluctuations on the electronic structure: Single-particle results}
\label{sec:electronic_structure_SP}

Figure~\ref{fig:SP_GS_electron_hole} shows the single-particle (a)
electron and (b) hole ground state (GS) energy as a function of the
microscopic configuration number (Config \#). The zero of energy for
all these calculations is the unstrained GaN valence band edge. The
data are shown for the different well width values ranging from
$L_w=1.6~$nm to $L_w=3.43~$nm. The results where WWFs have been
considered in the calculations are given by the filled symbols,
while the open symbols display the data in the absence of WWFs (No
WWF). Overall, two important features should be taken into account
before starting the detailed analysis of
Fig.~\ref{fig:SP_GS_electron_hole}. First, due to the presence of
the electrostatic built-in field along the growth direction (cf.
Fig.~\ref{fig:slice_phi_p}), electron and hole wave functions are
localized at opposite interfaces of the QW: electrons at the upper
interface and holes at the lower interface.~\cite{WaGo2011} Second,
it is important to remember that WWFs are introduced at the upper QW
interface. Thus, from a structural point of view, the lower
interface for a given microscopic configuration is unchanged. From
this perspective mainly electron GS energies should be affected by
WWFs. Equipped with this information, several general features can
be deduced from Fig.~\ref{fig:SP_GS_electron_hole}.

For both electrons and holes, the GS energies vary significantly
with configuration number, though the holes show a much greater
variance. This indicates that the alloy microstructure of the
systems plays an important role. Additionally, for all widths,
$L_w$, the inclusion of the WWF shifts the electron GS energies to
lower energies (cf. Fig.~\ref{fig:SP_GS_electron_hole} (a)).
Similarly for the hole GS energies (cf.
Fig.~\ref{fig:SP_GS_electron_hole} (b)), for all configurations but
one (Config 18), the inclusion of the WWF leads to a shift to higher
energies. We will come back to the particularities of Config. 18
further below.

For the electrons, Fig.~\ref{fig:SP_GS_electron_hole} (a), we
attribute the decrease in the GS energies to the slight increase in
QW volume due to the WWFs and the overall slightly stronger built-in
field in the system with WWFs when compared to structures without a
WWF (cf. Sec.~\ref{sec:Built_in_potential}). Turning to the hole GS
energy, even though the alloy microstructure at the lower interface
is not changed by the addition of a WWF, the changes in the QW
volume and the connected changes in built-in field due to the
presence of the WWF (cf. Sec.~\ref{sec:Built_in_potential}) also
affect the hole GS energy. However, when considering a given
configuration, the changes in the hole ground state energies due to
the presence of a WWF are in general smaller when compared to the
changes in the electron GS energies.

\begin{figure}
\includegraphics[width=\columnwidth]{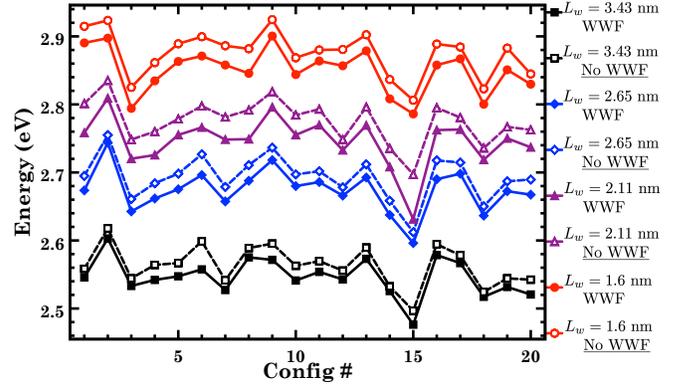}
\caption{Single particle transition energies as a function of the
microscopic configuration number (Config \#). The results are given
for the different well widths $L_w$. Data in the absence of the well
width fluctuation (No WWF) are denoted by the open symbols, while
data in the presence of the well width fluctuation (WWF) are given
by the filled symbols.} \label{fig:SP_GS_transition_energies}
\end{figure}

The resulting changes in the electron and hole GS energies due to
the presence of WWFs are summarized in
Fig.~\ref{fig:SP_GS_transition_energies}. Here the GS
\emph{transition energy} for the different well widths is displayed
as a function of the microscopic configuration number (Config \#).
For all well widths $L_w$ we observe a red shift in the transition
energy due to the presence of WWFs. This behavior reflects the
trends in the electron and hole GS energies. This finding is also
consistent with the simple picture that by including the WWF the
well becomes effectively wider and that the built-in field slightly
increases when compared to the QW without a WWF. Additionally, as
expected from our analysis above, the GS transition energy varies
significantly as a function of the configuration number. This
behavior is consistent with the experimentally observed broad PL
spectra in InGaN/GaN QWs.~\cite{GrSo05,ChAb98,DoMa99}

Having established general features of how alloy fluctuations and
WWFs affect both electron and hole GS energies in a single-particle
picture, we turn now and look at excitonic effects. Initially, we
present an analysis of the (average) transition energy and also the
exciton binding energy. In a second step we discuss how the Coulomb
effect impacts carrier localization features.
\begin{figure}[t!]
\includegraphics{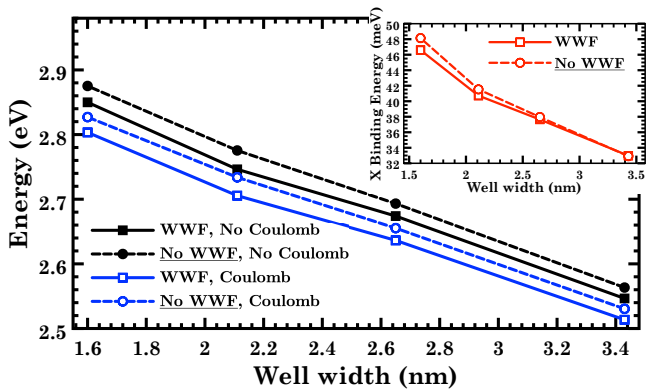}
\caption{Average transition energies with (WWF, squares) and without
(No WWFs, circles) well width fluctuations as a function of the well
width $L_w$. The data in the absence of Coulomb effects are given by
the respective filled symbols while the open symbols denote the
results when excitonic effects are taken into account. Inset:
Exciton (X) binding energies in the presence (WWF, open squares) and
absence (No WWF, open circles) of well width fluctuations as a
function of the well width $L_w$.} \label{fig:Tran_Energies_X}
\end{figure}

\subsection{Impact of well width, random alloy and well width fluctuations on the electronic and optical properties: Excitonic effects}
\label{sec:electronic_structure_MB}

Figure~\ref{fig:Tran_Energies_X} displays the \emph{average} GS
transition energies in the presence and absence of WWFs as a
function of the well width $L_w$. Additionally, results with and
without the inclusion of Coulomb effects in the calculations are
given. The open (filled) symbols denote the results in the presence
(absence) of excitonic effects. The squares (circles) show the data
when WWFs are included (excludes) in the calculations. Thus, the
filled symbols in Fig.~\ref{fig:Tran_Energies_X} are obtained from
averaging over the 20 different microscopic configurations presented
in Fig.~\ref{fig:SP_GS_transition_energies}. The filled symbols in
Fig.~\ref{fig:Tran_Energies_X} confirm in a simple manner the
aforementioned conclusion that the inclusion of WWFs result in a red
shift of the transition energies. Figure~\ref{fig:Tran_Energies_X}
also illustrates clearly the increase in transition energy with
decreasing $L_{w}$. Such a behavior is expected from a simple
particle-in-box-like picture. Looking at the open symbols, we see
that these expected trends also hold in the case where Coulomb
effects are included.

Additionally, a comparison of the average transition energy with and
without Coulomb effects allows for a first insight into the question
of how the Coulomb interaction affects the optical properties of QWs
with different structural parameters. To facilitate this
investigation, the inset in Fig.~\ref{fig:Tran_Energies_X} depicts
the exciton (X) binding energy in the presence (square) and abesence
(circle) WWFs, as a function of $L_{w}$. Here the exciton binding
energy is defined as the difference in the average single particle
GS transition energy and the calculated average excitonic transition
energy. Three important features can be extracted from this inset:
(i), the exciton binding energy increases with decreasing $L_{w}$;
(ii), in general the exciton binding energy decreases with inclusion
of WWFs; and (iii), the impact of the WWF on the excitonic binding
energy decreases with increasing $L_{w}$ and turns out to be of
secondary importance for large $L_{w}$. These results can be
explained in terms of wave function overlaps: in a simplified
picture one expects that the lower the spatial wave function
overlap, the lower the excitonic binding energy.
Thus, the decreasing vertical spatial separation of
electron and hole wave functions with decreasing $L_{w}$ accounts
for the increase in excitonic binding.
Secondly, if the electron wave functions are localized independently
of $L_w$ by the WWFs, then a further in-plane spatial separation, in
addition to the out-of plane separation by the electrostatic
built-in field, is expected. This in-plane separation can then lead
to a further decrease in the excitonic binding energy when comparing
the system with and without WWFs.
Finally, the finding that removing the WWF has a reduced effect on
the binding energies in wider than in narrower wells, is due to the
fact that the strength of the Coulomb interaction is determined
primarily by the vertical separation between the carriers.
The fact that for larger values of $L_w$ ($L_w > 2.5$ nm) the
average exciton binding energy is almost unaffected by the WWF,
leads to the following interesting conclusion. Assuming that WWFs
lead to an additional  in-plane spatial separation between the
carriers compared to a system without WWFs, the results of
Fig.~\ref{fig:Tran_Energies_X} indicate that the in-plane separation
between the carriers is of secondary importance, at least in terms
of the exciton binding energy. We will come back to this in
Sec.~\ref{sec:comp_exp}, when we compare our theoretical data with
experimental results and connect our theoretical findings also to
radiative recombination characteristics. It should be noted that
additional factors, such as the density and the statistical
variations in size and shape of WWFs, could affect this result.
These factors could further increase the in-plane spatial separation
contribution.

\begin{figure}[t!]
\includegraphics[width=\columnwidth]{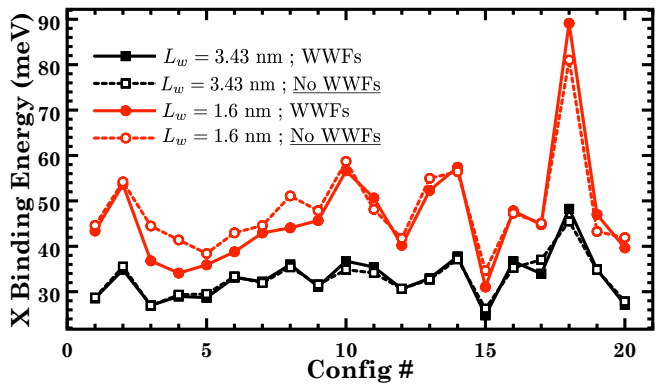}
\caption{Exciton (X) binding energy as a function of the microscopic
configuration number (Config \#). The results are shown for a well
width of $L_w=1.6$ nm (red circles) and $L_w=3.43$ nm (black
squares), respectively. The open symbols denote the results in the
absence of well width fluctuations (No WWFs) while the filled
symbols denote the data when well width fluctuations (WWFs) are
taken into account.} \label{fig:X_bind_WWF_NoWWF}
\end{figure}

So far we have only discussed \emph{averaged} results and made
assumptions about the nature of the carrier localization features.
To corroborate these explanations of observed trends in the averaged
properties and to investigate the interplay between well width,
alloy microstructure, structural inhomogeneities and Coulomb
effects, we present next an analysis of the electronic structure of
the different QWs for different microscopic configurations.
Figure~\ref{fig:X_bind_WWF_NoWWF} displays the exciton binding
energies as a function of the configuration number (Config \#). Here
we have focused our attention on the extreme well width cases, thus
$L_{w}=1.6~$nm (red circles) and $L_{w}=3.43~$nm (black squares).
The other well width systems considered here reflect then an
intermediate situation. The results in the absence of WWFs (No WWFs)
are always indicated by open symbols (square or circle) while the
results where WWFs are taken into account are denoted by the filled
symbols. We start our analysis here by looking at the QW system with
$L_w=3.43~$nm (black squares). When comparing the exciton binding
energies obtained in the presence and absence of WWFs, we find that
the WWF is of secondary importance for this quantity. This finding
is consistent with the results of the average exciton binding energy
depicted in the inset of Fig.~\ref{fig:Tran_Energies_X}. This
indicates already that for wider QWs, the wave function overlap is
not significantly affected by the presence of WWFs. Again, we will
come back to this question below when we discuss the interplay of
WWFs and Coulomb effects on the wave functions/charge densities. For
the system with $L_w=1.6$ nm, the results are given by the (red)
circles in Fig.~\ref{fig:X_bind_WWF_NoWWF}. In contrast to the
system with $L_w=3.43~$nm, we observe that the presence of WWFs can
make a noticeable difference to the exciton binding energy in the
different configurations. However, there is not a universal trend in
the sense that WWFs always lead to an decrease in the exciton
binding energy. We have also situations where the WWF leads to an
increase in the exciton binding energy (e.g. Configs. 18 and 19).
Overall, this reveals that for smaller $L_{w}$ values, the presence
or absence of WWFs in combination with Coulomb effects significantly
affect the electron-hole overlap. For larger well width, this seems
not to be the case since the exciton binding energies are almost
unaffected by the presence of WWFs compared to the situation where
WWFs are absent (cf. Fig.~\ref{fig:X_bind_WWF_NoWWF}).

\begin{figure*}
\includegraphics{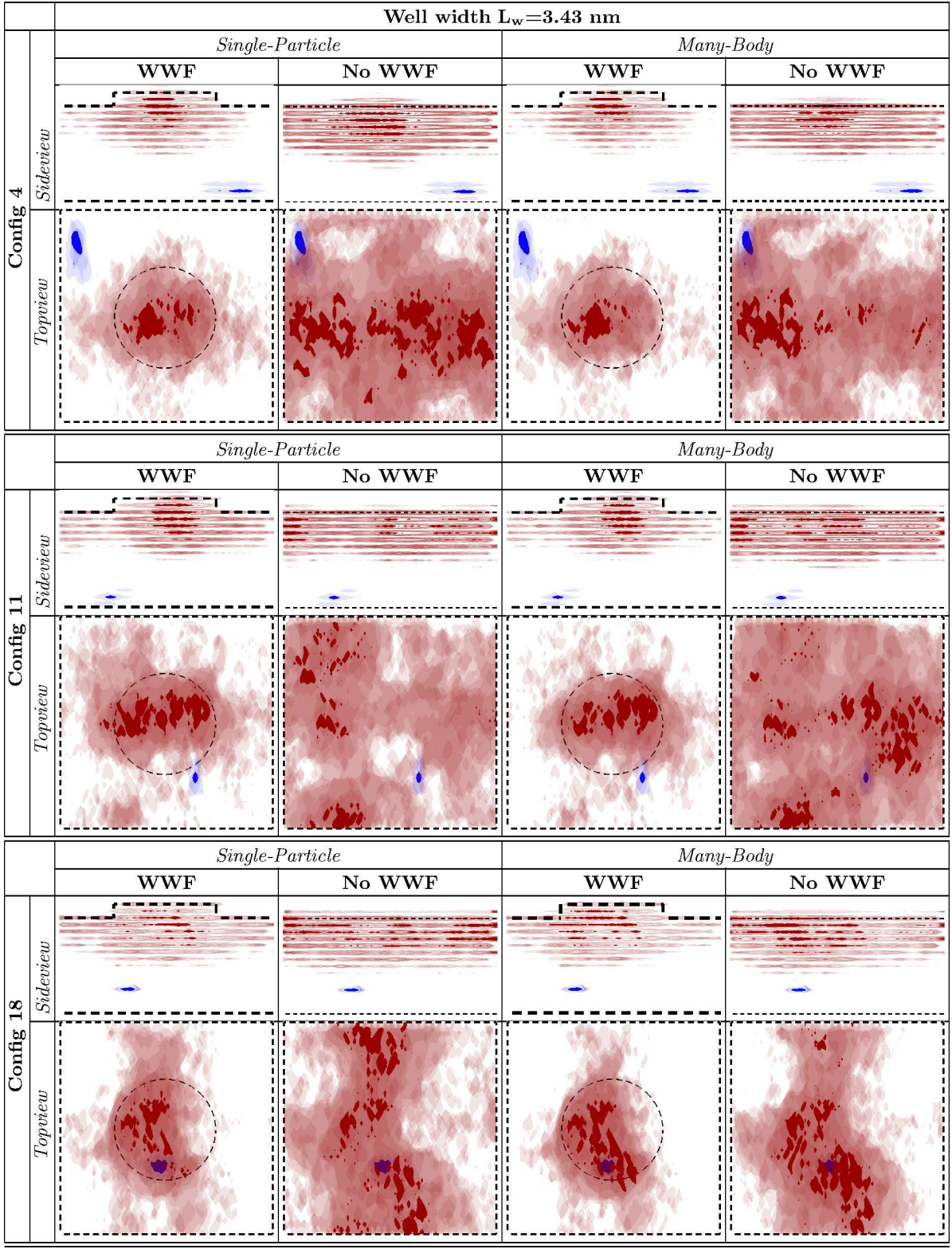}
\caption{Isosurface plots of electron (red) and hole (blue) ground
state charge densities in the absence (Single-Particle; left) and
presence (Many-Body; right) of Coulomb (excitonic) effects for
different microscopic configurations and different view points
(Sideview: Perpendicular to $c$-axis; Topview: Parallel to
$c$-axis). The light (dark) isosurfaces correspond to 10\% (50\%) of
the maximum charge density. Results are shown with (WWF) and without
(No WWF) well width fluctuations. The well width $L_w$ here is
$L_w=3.43$ nm.} \label{fig:Charge_Densities_343nm}
\end{figure*}

\begin{figure*}
\includegraphics{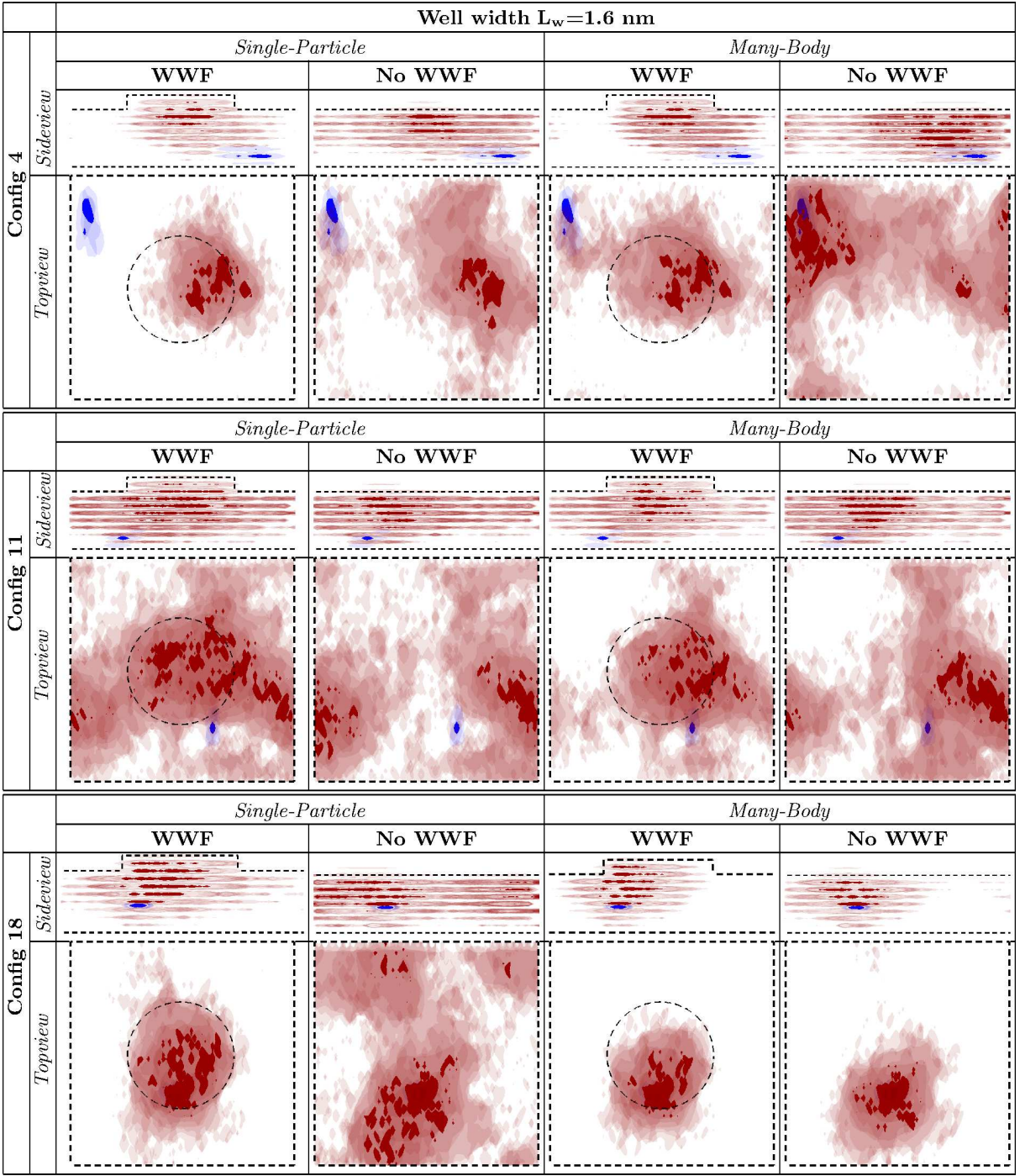}
\caption{Isosurface plots of electron (red) and hole (blue) ground
state charge densities in the absence (Single-Particle; left) and
presence (Many-Body; right) of Coulomb (excitonic) effects for
different microscopic configurations and different view points
(Sideview: Perpendicular to $c$-axis; Topview: Parallel to
$c$-axis). The light (dark) isosurfaces correspond to 10\% (50\%) of
the maximum charge density. Results are shown with (WWF) and without
(No WWF) well width fluctuations. The well width $L_w$ here is
$L_w=1.6$ nm.} \label{fig:Charge_Densities_16nm}
\end{figure*}

To shed more light onto wave function localization features and to
support explanations given earlier in terms of wave function
overlaps, we now turn to examine the single-particle and many-body
(excitonic) charge densities of the considered QWs.
Figures~\ref{fig:Charge_Densities_343nm}
and~\ref{fig:Charge_Densities_16nm} display isosurface plots of the
electron and hole GS charge densities in the presence and absence of
WWFs for the QWs with $L_w=3.43$ nm and $L_w=1.6$ nm, respectively.
Here, the results are displayed both for the single-particle states
(Single-Particle) as well as when Coulomb interaction effects
(Many-Body) are taken into account. The excitonic charge densities
have been deduced from reduced density
matrices.~\cite{BaSc2013,ScCa2015} For both systems, the results are
shown for a side view (perpendicular to $c$-axis) and for a top view
(parallel to the $c$-axis). The isosurfaces of the charge densities
are displayed at 10\% (light surface) and 50\% (dark surface) of the
respective maximum probability densities. Electron charge densities
are given in red, hole charge densities in blue. To understand
general features, we have selected configurations (Configs.) 4, 11
and 18. These configurations have been selected on the basis of the
system with $L_w=1.6$ nm since they reflect situations where the WWF
leads to a reduction (Config. 4) and an increase (Config. 18) of the
exciton binding energy (cf. Fig.~\ref{fig:X_bind_WWF_NoWWF}). The
last configuration, Config. 11, represents the situation where the
exciton binding is almost unaffected by the presence of WWFs (cf.
Fig.~\ref{fig:X_bind_WWF_NoWWF}).

\begin{figure}[t!]
\includegraphics{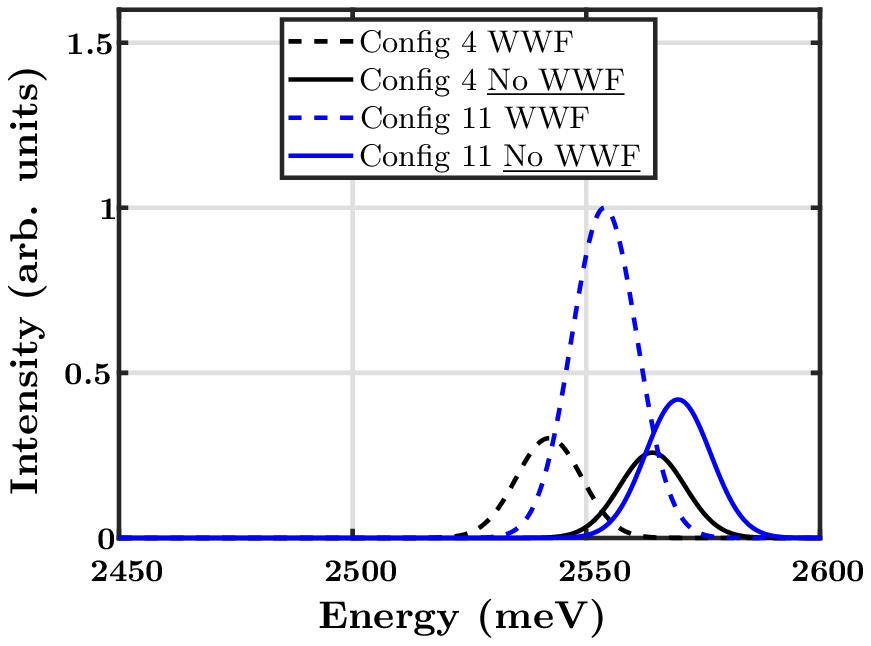}
\caption{(Color online) Emission spectrum, neglecting Coulomb
effects (single-particle), for Configs. 4 (black) and 11 (blue) in
the presence (dashed lines) and absence (solid lines) of well width
fluctuations. The well width of the system is $L_w=3.43$ nm. The
data is normalized to Config. 11 in the presence of well width
fluctuations (WWF).} \label{fig:Oscill_3_43_SP}
\end{figure}

We start our analysis with the QW system of width $L_w=3.43~$nm. The
charge densities of the above discussed configurations are displayed
in Fig.~\ref{fig:Charge_Densities_343nm}. In a first step we look at
the single-particle results with and without WWFs (left two columns
in Fig.~\ref{fig:Charge_Densities_343nm}). Firstly, due to the
presence of the electrostatic built-in field, electron and hole wave
functions are separated to opposite interfaces of the QW. Consistent
with previous calculations,~\cite{WaGo2011,ScCa2015,MaPe16,JoTe2017}
we find here very strong hole localization effects due to random
alloy fluctuations. Keeping in mind that for a given microstructure
(configuration) the local alloy arrangement at the lower interface
is not changed when including the WWF, the hole wave function
remains localized in the same spatial position when comparing the
results in the presence and absence of the WWF for the same
configuration. However, we observe that the localization features of
the electron wave functions, for all configurations, are changed due
to the presence of the WWF. We find that the electron wave function
is always localized at the upper interface by the WWF. This is
consistent with previous theoretical
studies.~\cite{Wats2011,WaGo2011,ScCa2015} In the absence of the
WWF, the electron wave function is more delocalized but still is
perturbed by the alloy microstructure. The origin of the anomalous
behavior of the hole GS energy of Config. 18, discussed in
Sec.~\ref{sec:electronic_structure_SP}, can readily be inferred from
Fig.~\ref{fig:Charge_Densities_343nm}. We find that for this
configuration, the hole GS is localized exceptionally high in the
QW. Therefore, the hole state will be significantly affected by
local changes in the built-in potential and strain field, introduced
by the presence of a WWF (cf. discussion in
Sec.~\ref{sec:Built_in_potential}). This extreme situation results
in the  feature observed in Fig.~\ref{fig:SP_GS_electron_hole}, that
Config. 18 exhibits a slight decrease in hole GS  energy with the
inclusion of the WWF, in contrast to all other configurations.
Furthermore, the spatial separation, both in- and out-of plane, of
electron and hole wave functions is unusually small. This leads to
an exceptionally large exciton binding energy (cf.
Fig.~\ref{fig:X_bind_WWF_NoWWF}).

To further elucidate the connection between carrier localization
features and wave function overlap, we have calculated the
single-particle emission spectrum for Configs. 4 and 11 in the
presence and absence of the WWF. The results are displayed in
Fig.~\ref{fig:Oscill_3_43_SP}. We do not present here the results
from Config.~18 since the oscillator strength in this case is more
than 20 times larger than the largest value of Config. 11. This
feature would distract here from the results of the more
``standard'' configurations. In the following all data are
normalized to the results from Config. 11 in the presence of the
WWF. For visualization purposes, each peak has been broadened by an
Gaussian. As we can see from Fig.~\ref{fig:Oscill_3_43_SP}, the
oscillator strength is approximately unchanged in case of Config. 4
when adding or removing the WWF. For Config. 11 the hole wave
function is localized near the WWF, explaining why in this case the
oscillator strength is higher when the WWF is present compared to
the case without the WWF (cf. Fig.~\ref{fig:Oscill_3_43_SP}).

\begin{figure}[t!]
\includegraphics{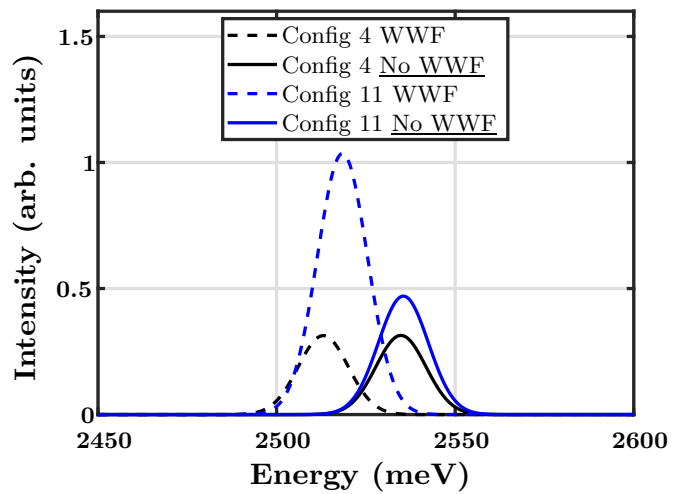}
\caption{(Color online) Excitonic emission spectrum for Configs. 4
(black) and 11 (blue) in the presence (dashed lines) and absence
(solid lines) of well width fluctuations. The well width of the
system is $L_w=3.43$ nm. The data is normalized to the spectrum of
Config. 11 in the absence of Coulomb effects but in the presence of
the well width fluctuation (WWF).} \label{fig:Oscill_3_43_MB}
\end{figure}

In a second step we study now how the (attractive) Coulomb
interaction affects the results. This data is shown on the right
hand side of Fig.~\ref{fig:Charge_Densities_343nm} (Many-Body) with
and without the WWF. When comparing these results with the
single-particle data, in the presence of the WWF, very little change
is observed in the spatial localization features of both electrons
and holes. This means that electron and holes are basically
``independently'' localized. For the case where there is no WWF, we
may still say that the electron and hole are independently localized
on opposite interfaces of the QW; however, more pronounced charge
density rearrangements for the electron are observed. This lateral
rearrangement of the electron charge density can most easily be seen
from the top view. In both cases (Config.~4 and~11), these
redistributions of the electron charge density about the hole wave
function should lead to a (slight) increase in the GS oscillator
strength. The calculated excitonic GS emission spectrum is shown in
Fig.~\ref{fig:Oscill_3_43_MB}. Indeed for Config. 11 and Config. 4,
the attractive Coulomb interaction leads in both cases, with and
without WWFs, to a slight increase in oscillator strength. Note here
that the data is again normalized to Config. 11 in the absence of
Coulomb effects but in the presence of the WWF (cf.
Fig.~\ref{fig:Oscill_3_43_SP}). The finding that with and without
WWFs we observe a only slight changes in the oscillator strength is
also consistent with the observation that for larger well width the
exciton binding energy does not change significantly (cf.
Fig.~\ref{fig:X_bind_WWF_NoWWF}) between the two systems (No WWF vs.
WWF).

\begin{figure}[t!]
\includegraphics{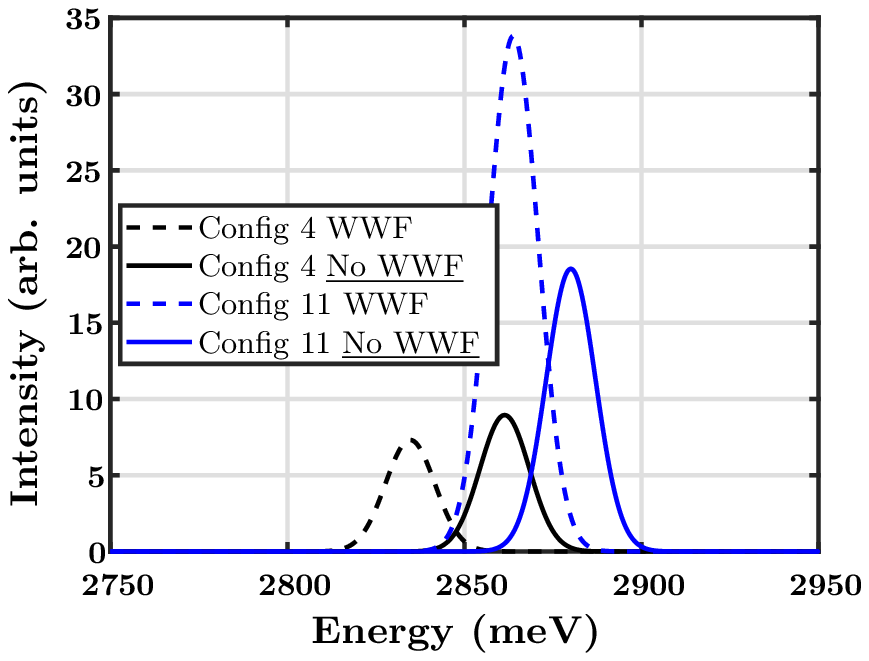}
\caption{(Color online) Emission spectrum, neglecting Coulomb
effects (single-particle), for Configs. 4 (black) and 11 (blue) in
the presence (dashed lines) and absence (solid lines) of well width
fluctuations. The well width of the system is $L_w=1.6$ nm. The data
is normalized to the spectrum of Config. 11 in the absence of
Coulomb effects but in the presence of the well width fluctuation
(WWF) (cf. Fig.~\ref{fig:Oscill_3_43_SP}).}
\label{fig:Oscill_1_6_SP}
\end{figure}

Taking these results together with the observations on the exciton
binding energies, Fig.~\ref{fig:X_bind_WWF_NoWWF}, one can draw the
following conclusion: For larger QWs, the excitonic binding energy
is basically independent of lateral localization features. This can
readily be inferred from Configs. 4 and 11 in
Fig.~\ref{fig:Charge_Densities_343nm}, where we observe that,
despite of the noticeable difference in in-plane separation between
the cases with and without WWFs, the differences in the binding
energies are negligible.
Furthermore, we find here that Config. 18 has the largest binding
energy (cf. Fig.~\ref{fig:X_bind_WWF_NoWWF}) of all configurations
due to the smallest spatial separation of electron and hole GS wave
functions.
Also from the comparison of Figs.~\ref{fig:Oscill_3_43_SP}
and~\ref{fig:Oscill_3_43_MB} we can deduce that the Coulomb effect
is of secondary importance for the wave function overlap, at least
for large well width. However, the data indicates already that in
general reducing the QW barrier interface roughness and therefore
WWFs should, even for these larger QW width, allow for a
redistribution of the (electron) charge density due to Coulomb
effects. This gives already indication that Coulomb effects can
reduce the spatial in-plane separation of the carriers. Overall, the
oscillator strength should at least slightly benefit from this
feature.

Having discussed the $L_w=3.43$ nm case, we now turn to the QW
system with $L_w=1.6$ nm. From Figs.~\ref{fig:Tran_Energies_X}
and~\ref{fig:X_bind_WWF_NoWWF} we know already that, compared to the
system with $L_w=3.43~$nm, the exciton binding energy is increased
in the $L_w$=1.6 nm structures on average. Also in
Fig.~\ref{fig:X_bind_WWF_NoWWF} we observe that both WWFs and the
microscopic alloy structure lead to a significant variation in the
exciton binding energy. To understand these differences, we look
first at the electron and hole GS charge densities. These are
presented in Fig.~\ref{fig:Charge_Densities_16nm}, using the same
configurations as in Fig.~\ref{fig:Charge_Densities_343nm}.

We begin our analysis, as before, by looking at the single-particle
properties (first two columns). Note that for each configuration,
the hole charge densities are essentially the same as in the
$L_w=3.43~$nm case, originating from the fact that for a given
configuration the alloy microstructure at the lower interface of the
well is unchanged when increasing the well width (cf.
Sec.~\ref{sec:QW_structures}). The GS electron charge densities,
being subject to a different atomic environment, show noticeable
differences from the QW with $L_w=3.43~$nm. However, the general
behavior is similar, with the electron being strongly affected by
the combined effect of built-in field and WWFs. Due to the smaller
QW width, electron and hole wave functions are less vertically
separated, leading to a higher spatial overlap between the wave
functions. This is also reflected in the calculated single-particle
GS emission spectrum shown in Fig.~\ref{fig:Oscill_1_6_SP}, again
for Configs. 4 and 11 with and without WWFs. As before, this data is
normalized to the spectrum of Config. 11 for the well with
$L_w=3.43$ nm in the absence of Coulomb effects but in the presence
of the WWF (cf. Fig.~\ref{fig:Oscill_3_43_SP}). Comparing the
results for a well width of $L_w=1.6$ nm
(Fig.~\ref{fig:Oscill_1_6_SP}) with data for $L_w=3.43$ nm
(Fig.~\ref{fig:Oscill_3_43_SP}), the oscillator strength increases
by a factor of order of at least 20. Otherwise, the general features
between the two systems are very similar. However, due to the
increased spatial electron and hole wave function overlap and
consistent with previous discussion, the exciton binding energy is
significantly increased for the well with $L_w=1.6$ nm when compared
to the $L_w=3.43$ nm system (cf. Fig.~\ref{fig:X_bind_WWF_NoWWF}).

\begin{figure}[t!]
\includegraphics{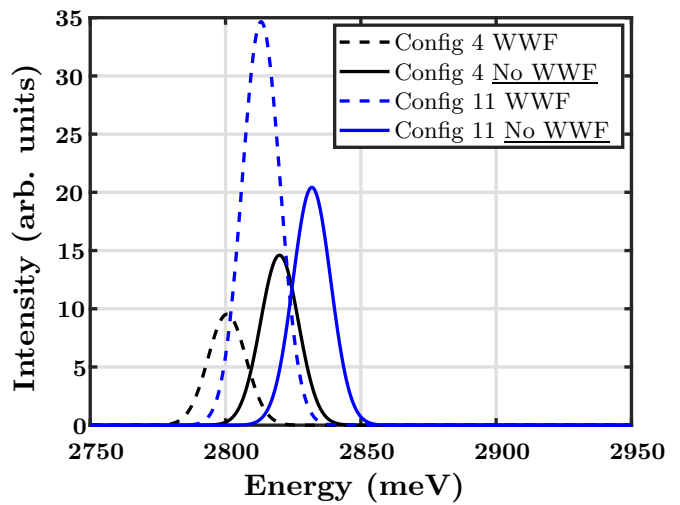}
\caption{(Color online) Excitonic emission spectrum for Configs. 4
(black) and 11 (blue) in the presence (dashed lines) and absence
(solid lines) of well width fluctuations. The well width of the
system is $L_w=1.6$ nm. The data is normalized to the spectrum of
Config. 11 in the absence of Coulomb effects but in the presence of
the well width fluctuation (WWF).} \label{fig:Oscill_1_6_MB}
\end{figure}

For the well with $L_w$=1.6 nm, a pertinent question now relates to
the balance between the enhanced Coulomb effect and the localization
``strength'' of WWF and random alloy: Is the Coulomb interaction
strong enough at this width to overcome the localizing potential of
the WWF and mitigate the lateral separation between the electron and
hole? To investigate this, we consider the charge densities obtained
in the presence of Coulomb effects. These results are shown in the
last two columns of Fig.~\ref{fig:Charge_Densities_16nm}
(Many-Body). Initially, we focus our attention on the situation with
WWFs. Looking at Config. 4, and comparing with the data from
$L_w=3.43$ nm (Fig.~\ref{fig:Charge_Densities_343nm}), we observe
clearly a stronger redistribution of the electron charge density
towards the hole. Similar findings hold for Configs. 11 and 18.
However, the carrier localization effects are still largely
determined by random alloy contributions and WWFs. In the presence
of WWFs, even for the lowest well width studied here and independent
of the microscopic configuration, the single-particle picture still
gives a very good approximation of the system in terms of its
localization features. Thus this system is still consistent with the
picture of ``independently'' localized carriers, as introduced by
Morel~\emph{et al.}~\cite{MoLe02} But, when looking at the
corresponding excitonic emission spectrum,
Fig.~\ref{fig:Oscill_1_6_MB}, and comparing to the single-particle
emission spectrum, Fig.~\ref{fig:Oscill_1_6_SP}, even in the
presence of WWFs (dashed lines) the oscillator strength is
noticeably affected by the attractive Coulomb effect. Note again,
the data is normalized to the spectrum of Config. 11 from
Fig.~\ref{fig:Oscill_3_43_SP}. We will come back to the impact of
the interplay of Coulomb effects and WWFs on the wave function
overlap in more detail below.


Taking the results from the $L_w=3.53$~nm and $L_w=1.6$~nm cases
together we are left with the following picture. Our calculations
indicate that the importance of the Coulomb interaction depends
strongly on the the vertical separation of the electron and hole
wave functions. Thus, in our case here on the the QW width. However,
a similar effect would be observed with increasing In content, given
that the strain dependent piezoelectric potential increases with
increasing In content and thus leads to a stronger out of plane
separation of the carrier wave functions.~\cite{WiSc2009} This
statement is supported by our recent studies, where we have shown
that with increasing In content carrier localization effects of
electrons and holes at the QW barrier interface are further enhanced
and the out-of plane carrier separation is
increased.~\cite{TaCa2016_RSC} Furthermore, we find here that WWFs
serve as a barrier to the lateral charge density redistribution
facilitated by Coulomb effects.

\begin{figure}[t]
\includegraphics{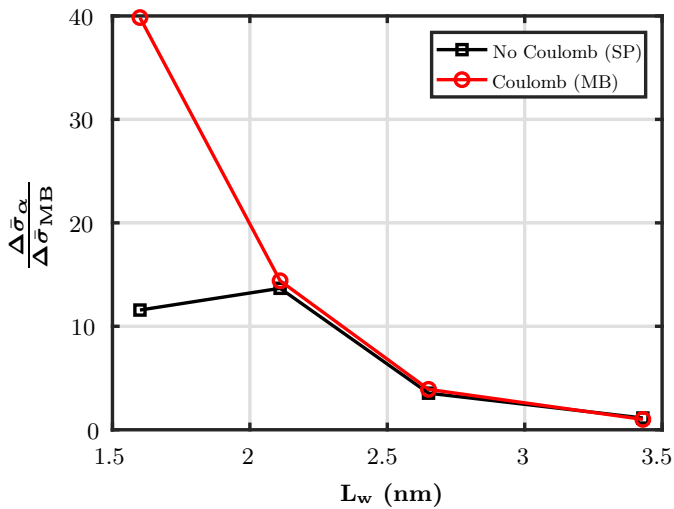}
\caption{(Color online) Normalized difference in the average peak
oscillator strength, $\Delta\bar{\sigma}_\alpha(L_w)$, between the
system without and with the WWFs as function of the well width
$L_w$. The results are shown in presence (red circle) and absence
(black squares) of Coulomb effects. More details are given in the
text.} \label{fig:Oscill_ratio}
\end{figure}

However the situation is less clear cut when WWFs are absent. For
the largest well width considered here, $L_w=3.43$ nm, charge
density redistributions are observed but for instance the oscillator
strength or the excitonic binding energies are only slightly
affected. But, when reducing the well width these charge
redistributions due to Coulomb effects are more pronounced and can
partially compensate the in-plane carrier separation.

This argument is supported by the data displayed in
Fig.~\ref{fig:Oscill_ratio},  which shows the (normalized)
difference in the average peak oscillator strength,
\mbox{$\frac{\Delta\bar{\sigma}_\alpha(L_w)}{\Delta\bar{\sigma}_\text{MB}(L_w=3.43
\text{
nm})}=\frac{(\bar{\sigma}^\text{NoWWF}_\alpha(L_w)-\bar{\sigma}^\text{WWF}_\alpha(L_w))}{(\bar{\sigma}^\text{NoWWF}_\text{MB}(L_w=3.43
\text{ nm})-\bar{\sigma}^\text{WWF}_\text{MB}(L_w=3.43 \text{
nm}))}$}, between the system without (No WWF) and with WWFs. The
results are displayed as a function of the well width $L_w$. The
index $\alpha$ denotes if the calculations have been performed in
the presence (MB, red circles) or absence (SP, black squares) of
Coulomb effects. Here, the data is normalized to
$\Delta\bar{\sigma}_\text{MB}(L_w=3.43 \text{ nm})$.
Figure~\ref{fig:Oscill_ratio} reveals that for the largest well
width removing the WWF has very little effect on the oscillator
strength \mbox{($\Delta\bar{\sigma}_\alpha(L_w=3.43 \text{
nm})/\Delta\bar{\sigma}_\text{MB}(L_w=3.43 \text{ nm})\approx1$)}.
However, with decreasing well width $L_w$,
\mbox{$\Delta\bar{\sigma}_\alpha/\Delta\bar{\sigma}_\text{MB}>>1$},
indicating that the oscillator strength in the absence of the WWF
increases on average much more quickly when compared to the
situation where the WWF is present. This finding holds also for
situation with and without Coulomb effects. Note, the kink in the
data without Coulomb effects (black squares) is caused by the
extreme configuration 18. When neglecting Config. 18 (not shown)
$\Delta\bar{\sigma}_\text{SP}(L_w=1.6 \text{
nm})/\Delta\bar{\sigma}_\text{MB}(L_w=3.43 \text{ nm})\approx16$ and
$\Delta\bar{\sigma}_\text{SP}(L_w=2.11 \text{
nm})/\Delta\bar{\sigma}_\text{MB}(L_w=3.43 \text{ nm})$ decrease
slightly. Overall our analysis strengthens the argument that
reducing the roughness of the QW barrier interface can have a
significant impact on the optical properties of InGaN/GaN QWs and
thus on devices utilizing these. We discuss these benefits in more
detail in the following section.

\subsection{Comparison with experiment and consequences for radiative recombination in InGaN/GaN QWs}
\label{sec:comp_exp}

Taking the results of all the above studies into account we can draw
here the following conclusions, which are relevant for the radiative
recombination dynamics and optical properties of InGaN/GaN QWs.
Looking at Fig.~\ref{fig:Oscill_ratio} we note here that in the
extreme case of low well width and no WWFs, Coulomb effects have a
very strong impact on
$\Delta\bar{\sigma}_\text{MB}(L_w)/\Delta\bar{\sigma}_\text{MB}(L_w=3.43
\text{ nm})$ reflecting the above discussed observation that in the
absence of WWFs the spatial in-plane separation can be reduced by
the Coulomb effect. This has finding has now two immediate
consequences.

First, if the green gap problem is related to in-plane carrier
localization effects, as argued in Refs.~\onlinecite{MaPe16}
and~\onlinecite{Karpov_17}, reducing the surface roughness of GaN on
InGaN might lead to a significant improvement in the device
performance, at least for the radiative recombination, given the
charge density redistribution due to Coulomb effects and the
connected reduction of the spatial in-plane separation. Our finding
here is inline with the recent experimental studies focussing on
this aspect.~\cite{ChWu2004,WuSh2016,MaPi2017} Our calculations
confirm and support the recent experimental arguments that
understanding and tailoring the QW barrier interface in terms of
roughness plays an important role for the electronic and optical
properties of these systems. Furthermore, our argument that the
interface roughness plays a key role for optical properties of these
systems might also be related to the experimentally reported
enhancement of the PL intensity of InGaN/GaN QWs after reduction of
the surface roughness via Hydrogen treatment during
growth.~\cite{ZhLu17}

In addition to these device related aspects, our above presented
results indicate also potential changes in the fundamental physical
properties of the radiative recombination dynamics of InGaN/GaN QWs
with varying well width. Here, we find for instance that for some
configurations the recombination process may be considered as being
driven by exciton localization effects (cf. Configs 4 and 18 in
Fig.~\ref{fig:Charge_Densities_16nm}), with the electron localizing
about the hole, instead of a picture of independently localized
carriers. Reducing the well width and/or the In content further and
thus the built-in field, might lead to the situation of exciton
localization in general, independent of the microscopic
configuration. Therefore, our results give indications that a
continuous transition from a system that can be described by
independently localized carriers to the case where radiative
recombination is determined by exciton localization effects may be
achieved. Experimentally such a transition could be observed by
changes in the time dependent PL decay curves. For instance a
non-single-exponential decay in $c$-plane InGaN/GaN QWs has been
often explained by the picture of independently localized
carrier.~\cite{MoLe02,DaSc16} For non-polar systems,
single-exponential decay curves have been observed and exciton
localization effects have been used to explain this
behavior.~\cite{MaKe2013,ScTa2015,DaSc16} Interestingly, such an
effect of a change in PL decay characteristics has been observed
experimentally by Davidson~\emph{et al.}~\cite{DaDa00} and
Langer~\emph{et al.}~\cite{LaCh13} in $c$-plane systems. In the
experiments of Davidson~\emph{et al.}~\cite{DaDa00} the well width
of $c$-plane InGaN/GaN QWs with 13\% In has been reduced from 5 nm
to 1.25 nm revealing the transition from non-exponential to
single-exponential decay curves. In the study of Langer~\emph{et
al.}~\cite{LaCh13}, a marked difference in the carrier dynamics of a
2.7 nm and 1.1 nm QW are observed, with the constancy of the PL
decay lifetime of the smaller well with carrier density imputed to
strong excitonic effects.

Furthermore, Langer~\emph{et al.}~\cite{LaCh13} as well as
Hangleiter \emph{et al.}~\cite{HaJi2015} discussed in their work
excitonic recombination at room temperature. To achieve such an
behavior large exciton binding energies, exceeding the room
temperature thermal energy of 26 meV are required. Exciton binding
energies of 15-50 meV have been reported in combined theoretical and
experimental literature studies on GaN and InGaN
QWs.~\cite{BiLe1999,LaHi2003} These values are in good agreement
with our obtained values of 30-50 meV (cf.
Fig.~\ref{fig:X_bind_WWF_NoWWF}) for In$_{0.15}$Ga$_{0.85}$N/GaN
QWs. Therefore, the here calculated exciton binding energies, which
exceed the thermal energy  at room temperature, might facilitate the
existence of stable excitons at room temperature as reported
experimentally.

\section{Conclusion}
\label{sec:Conclusion}

In summary, we have presented a detailed theoretical analysis of the
impact of the interplay between well width, structural
inhomogeneities, alloy fluctuations, and Coulomb effects on the
electronic and optical properties of InGaN/GaN QWs. This analysis
not only gave insight into fundamental properties of this material
system, it also revealed possible routes towards the optimization of
the radiative recombination process in InGaN-based devices via
tailoring their structural properties.

Overall our calculations reveal that for the here studied QW widths
$L_w$, ranging from $L_w=1.6$ nm up to $L_w=3.43$ nm, carrier
localization effects due to built-in field, well width fluctuations
(electrons) and random alloy fluctuations (holes) dominate over the
attractive Coulomb interaction between the carriers. At least in
terms of wave function localization features, we are left with the
picture of ``independently'' localized carriers, consistent with the
concept introduced by Morel~\emph{et al.}~\cite{MoLe02} to explain
time-resolved PL spectra $c$-plane InGaN/GaN QWs.

Our data also shows that if well width fluctuations are present, in
addition to the built-in field induced out-of plane wave function
separation, an additional in-plane component is present. This
component is only slightly affected when Coulomb effects are taken
into account. Auf der Maur~\emph{et al.}~\cite{MaPe16} argued that
this in-plane separation feature is a key component to the ``green
gap'', though their calculations neglected well width fluctuations
and Coulomb effects.

However, our study shows that in the absence of structural
inhomogeneities such as well width fluctuations the situation is
less clear cut. In this case, and neglecting Coulomb effects
initially, carrier localization effects are introduced by the
interplay of random alloy fluctuations and built-in field. Here, we
find that hole wave functions are strongly localized by the random
alloy fluctuations while the electron wave functions are more
delocalized when compared to the situation with a well width
fluctuation. But, the electron wave function, in terms of
localization features, is perturbed by the random alloy
fluctuations. Thus, this gives rise to an in-plane spatial
separation of electron and hole wave functions. When including
Coulomb effects in the calculations, for larger well width, only
slight modifications of the charge densities are observed. However,
these results indicated a charge density redistribution of the
electron wave function, while the hole charge density is basically
unchanged. This feature is more pronounced with decreasing well
width, clearly revealing that in the absence of well width
fluctuations, the in-plane spatial separation of electron and hole
wave functions in a single-particle picture can be reduced by the
attractive Coulomb interaction between the carriers.

Therefore, our analysis gives the important finding that reducing
the GaN InGaN interface roughness in InGaN/GaN QWs should be
extremely beneficial for the radiative recombination in light
emitters utilizing this material system. This originates from the
observation that in the absence of well width fluctuations the
in-plane spatial carrier localization can be partially compensated
by the attractive Coulomb interaction.

While these features are of interest from a device application point
of view, especially for the green gap problem, our results also
indicate that it might be possible to ``tune the physics'' of this
system by changing the well width. For instance, we observe here
that for the smallest well width ($L_w=1.6$ nm), for some
configurations the electron localizes about the hole due to the
attractive Coulomb interaction. Thus we observe here exciton
localization features in $c$-plane structures instead of the picture
of ``independently'' localized carriers usually used to describe the
optical properties of these system. Experimentally such a transition
could be observed in time resolved PL measurements for instance and
has indeed been observed in $c$-plane systems.~\cite{DaDa00,LaCh13}
Moreover, our here presented analysis, in terms of carrier
localization and impact of Coulomb effects should also be of
interest for studies on novel monolayer InGaN/GaN
systems.~\cite{MaRo2017}

\begin{acknowledgements}
This  work  was  supported  by  Science  Foundation  Ireland
(project number 13/SIRG/2210).
\end{acknowledgements}

\bibliography{./Daniel_Tanner_Bibliography.bib}

\end{document}